\documentclass[a4paper]{jpconf}
\usepackage{graphicx}
\begin{document}
\title{Information and Cosmological Physics}

\author{Rajesh R. Parwani}

\address{Department of Physics, National University of Singapore, Kent Ridge, Singapore.}

\ead{parwani@nus.edu.sg}

\begin{abstract}
We review an information-theoretic approach to quantum cosmology, summarising the key results obtained to date, including a suggestion that an accelerating universe will eventually turn around.  
\end{abstract}

\section{Preamble}
There are several intriguing approaches to quantum spacetime, and quantum cosmology in particular, as evidenced by the talks at this conference and the papers on arXiv which the reader may consult. 

Here we will describe the work of a number of people involved in one endeavour: An information-theoretic approach to quantum cosmology. Spacetime limits will make this review brief and qualitative: We invite the interested reader to consult the cited papers for details and more references.

\section{A Method of Inference}

The maximum entropy principle (MEP) \cite{Jaynes} originated in statistical mechanics as an inference principle which allowed one to obtain relevant probability distributions in a conceptually appealing manner. For example, consider a statistical system with unknown probability distribution $p(x)$ but with specified mean energy $E = \int \epsilon (x) \ p(x) \ dx$.  In the MEP approach one maximises the Gibbs entropy 
\begin{equation}
I_{GS} = -\int p(x) \ln p(x) \ dx \label{gs}
\end{equation}
under the given constraint to determine the form for the probability distribution: Introducing the Lagrange multiplier $\beta$ and maximising $I_{GS} - \beta E$ with respect to variations in $p(x)$ gives the well known canonical probability distribution $p(x) \propto \exp(-\beta \epsilon (x))$. 

Although in physics the quantity (\ref{gs}) is usually associated with problems in statistical mechanics, an identical expression  was derived independently by Claude Shannon in his search for a measure that could be used to quantify the information content, or uncertainty, in a system.  
Indeed the original MEP of physics is actually an example of a more general maximum uncertainty principle (MUP), a method of inference used in diverse fields of study \cite{apply}: The basic idea is that one should provide the most unbiased description of the state of the system, since maximising the uncertainty  measure acknowledges our ignorance of a more detailed structure.

In some cases one may already have some {\it a priori} information about the system, encoded in the form of a reference probability distribution $r(x)$. Then  the relevant measure that is maximised is a relative uncertainty measure called the Kullback-Leibler (KL) information,  
\begin{equation}
I_{KL}(p,r) = -\int p(x) \ln {p(x) \over r(x)} \ dx \label{kl}
\end{equation}
If there is no useful {\it a priori} information then $r(x)$ can be taken to be a uniform distribution and the Kullback-Liebler measure then reduces to the Gibbs-Shannon entropy.

Although the MUP is elegant, it does require some specific input: The precise form of the information measure to be used. As Shannon showed, the measure (\ref{gs}) is the simplest measure that satisfies certain axioms appropriate for the context of some problems. Different situations may require one to relax those assumptions and thus lead to measures such as (\ref{kl}) or others like the Fisher measure to be discussed below.

\section{Quantum Physics}
It is useful to re-examine the Schrodinger equation, which describes the evolution of probability amplitudes, from the perspective of the MUP \cite{sch}.
In one dimension the transformation $\psi = \sqrt{p} \ e^{iS/ \hbar}$ can be used to re-write the Schrodinger equation in terms of two real functions, $p$ and $S$, 
\begin{eqnarray}
{\partial S \over \partial t} + {1 \over 2m} \left( {\partial S \over \partial x} \right)^2 + V + Q &=& 0 \, ,  \label{hj1} \\
{\partial p \over \partial t} + {1 \over m} {\partial \over \partial x} \left(p {\partial S \over \partial x} \right)&=&0  \, ,   \label{cont1}
\end{eqnarray}
with $Q =  - {{\hbar}^2 \over 2m}  {1 \over \sqrt{p}} {\partial^2 \sqrt{p} \over \partial x^2}$ the ``quantum potential".  For $Q=0$, the  equation (\ref{hj1}) is just the Hamilton-Jacobi equation for a classical {\it ensemble} of particles described by a probability distribution $p(x,t)$ and with the function $S(x,t)$ related to the velocity of a particle by $v= {1 \over m} {\partial S \over \partial x}$. The second equation (\ref{cont1}) is just the expression for conservation of probability.

It was noted \cite{sch} that the classical, $Q=0$, limit of equations (\ref{hj1},\ref{cont1}) may be obtained by minimising the action 
$ \Phi_A \equiv \int p \left( {\partial S \over \partial t} + {1 \over 2m} \left( {\partial S \over \partial x} \right)^2 + V \right)  dx \ dt $
through a variation of both $p$ and $S$. The full quantum equations follow if simultaneously the Fisher information (an inverse uncertainty measure)
$ I_F = \int {1 \over p} \left( {\partial p \over \partial x} \right)^2  dx \ dt $ 
is also minimised: That is, $\Phi_A + \xi I_F $ is minimised with respect to both $p$ and $S$. If the Lagrange multiplier $\xi$  is set to  ${\hbar}^2 / 8m$ then one obtains the equations (\ref{hj1},\ref{cont1})  which are equivalent to the time-dependent Schrodinger equation. 

Thus the principle of minimum Fisher information is an example of the MUP and the result above may be viewed as an extension of the method of inference to quantum mechanics.  More specifically, one may argue that the method of inference allows one to supplement classical ensemble dynamics with additional fluctuations to arrive at quantum mechanics. The method is easily extended to many particles in higher dimensions.

\section{Nonlinear Schrodinger Equation}

The pioneering approach of \cite{sch} did not provide a derivation of the Fisher measure in the same way that the Gibbs-Shannon measure was derived: From axioms appropriate for the information-theoretic context. Such a derivation was provided in \cite{fish}, and it clarified that the Fisher measure was the simplest measure to satisfy suitable assumptions such as locality and separability. One of the assumptions \cite{fish} was that the number of derivatives in the information measure be at a minimum. Clearly then, a generalisation would lead to higher derivative information measures and a generalised Schrodinger equation.

However a direct expansion in derivatives would likely lead to highly singular expressions in the generalised Schrodinger equation and so a different approach would be to construct a suitable information measure which may be viewed as a ``sum of higher derivative" terms. 
It turns out that there already exists such a measure!  If in eq.(\ref{kl}) one chooses the reference  distribution $r(x)$ to be the same as $p(x)$ but with infinitesimally shifted arguments, that is $r(x) = p(x + \Delta  x)$, then to lowest order,
\begin{eqnarray}
I_{KL} ( p(x), p(x + \Delta(x)) &=& {- (\Delta x)^2  \over 2} I_F (p(x)) + O(\Delta x)^3  \label{expan}
\end{eqnarray}
So to lowest order, minimising the Fisher information is the same as maximising the relative entropy for two probability distributions that are  close to each other, $r(x) = p(x + \Delta x)$. 

An interpretation of $\Delta x$ is readily available: For example, many heuristic arguments combining quantum theory and gravity suggest the existence of a minimal position uncertainty, thus $\Delta x$ in (\ref{expan}) may be taken to be the scale at which the coordinates become distinguishable. 
Beyond leading order, the expansion in (\ref{expan}) involves  higher derivatives. These too have a natural interpretation as describing situations where fluctuations at increasingly shorter scales become important, as one might expect when quantum physics affects spacetime.   
 
This motivates one to study a generalised Schrodinger equation that results from using the KL measure instead of the Fisher measure in the derivation of the last section. It leads to a nonlinear Schrodinger equation \cite{RP1} whose consequences have been studied in a number of papers.   Though nonlinear, the equation still shares some important properties of the linear Schrodinger equation: The continuity equation (\ref{cont1}) is unchanged so that $p=\psi^{*} \psi$ is still conserved and has a sensible interpretation as probability density; also the equation does not depend on the normalisation of the wavefunction. 

A conservative attitude would be to view the information-theoretic nonlinear Schrodinger equation as an effective equation which models, perhaps in an approximate manner, the unknown.

In passing we note that nonlinear generalisations of the Dirac equation using information-theoretic measures have been discussed in \cite{wk}.

\section{Quantum Cosmology of de Sitter Space} 
In quantum cosmology one studies the universe as a single quantum entity, which in the Wheeler-DeWitt (WDW) approach is described by a wavefunction, the wavefunction of the universe.  Usually situations of restricted symmetry are studied, the minisuperspace scheme, which reduces the functional differential equation to a manageable Schrodinger-like equation with few degrees of freedom. 

In the spirit of the MUP, it was suggested in \cite{nguyen} that an information-theoretic extension of the usual WDW equation be used to model the unknown structure of quantum spacetime. This was motivated in part by the fact that the usual linear WDW equation could not always resolve cosmological singularities.

 The nonlinear WDW equation for de Sitter space studied in \cite{nguyen} was just the nonlinear Schrodinger equation of \cite{RP1} but using the potential appropriate for quantum cosmology. As the equation was complicated, being a nonlinear difference-differential equation, the first attempt involved studying a  truncated and linearised version, with an effective potential which represented approximately the new ingredients.
 Encouraging results were obtained. It was shown that the original Big Bang of the linear WDW equation was replaced by a creation of the universe through tunneling. 
 
 Recently \cite{siti}, we managed to transform the full nonlinear difference-differential equation into a purely difference equation for the probability density by solving first the current conservation constraint. The difference equation could then be studied easily, though still mostly numerically. The full treatment showed some new features not seen in the previous approximate study: The existence of a minimum and maximum allowable size to the quantum de Sitter universe. 
 
At the quantum mechanical level, $a_{min}$ and $a_{max}$ corresponded to the location where the probability density $p$ vanished. Unlike the case for normal quantum mechanical systems, the vanishing of $p$ in the nonlinear WDW equation with a de Sitter potential was shown to imply a termination of the evolution of the difference equation. 

Since for the de Sitter case there is no other variable in the WDW equation to play the role of an internal clock, a semi-classical analysis was performed to interpret $a_{min}$ and $a_{max}$. It was shown that in the effective classical dynamics the location of nodes in $p$ corresponded to the position of barriers which caused bounces at short and large distances.

The de Sitter model is reasonable at early times of our universe. It will also be a reasonable model at late times if the current acceleration continues: However our results, taken at face value, suggest that eventually the acceleration will stop and the universe collapse.   

The results above hold even when the cosmological constant varies slowly \cite{meer}.

\section{The Quantum FRW-$\phi$ Universe}

A massless scalar field may be used as an internal clock. In our first approximate study in \cite{nguyen}, using the truncated  version of the nonlinear WDW equation, we found that the zero size singularity of the classical FRW-$\phi$ model was resolved by a bounce in the effective classical dynamics. 

Recently we \cite{meer}  extended the study of the FRW-$\phi$ model by using the full nonlinear difference-differential equation in a similar manner to \cite{siti}. By using a suitable ansatz, wavepackets with appropriate physical properties were constructed. It was observed that the quantum evolution of the wavepackets displayed bounces at short and large distances leading to cyclic evolution, though the cycles observed so far are not periodic nor everlasting.

\section{Outlook}

The results from our studies of toy quantum universes, within the information-theoretic approach, have been encouraging. Some natural extensions would involve adding other matter fields, studying less symmetrical situations, and examining the robustness of the results to deformations of the information measure used. 

\ack
The quantum cosmology results reported here were obtained in collaboration with L.H. Nguyen, S.N. Tarih and M. Ashwinkumar. 
R.P thanks the organisers of ICGC 2011 (Goa, India) for the opportunity to present this summary.


\section*{References}

\begin{thebibliography}{99}



\bibitem{Jaynes} Jaynes E T 1957 {\it Phys. Rev.} {\bf 106}  620 

\bibitem{apply} Jaynes E T 2004  \textit{Probability Theory, The Logic of Science} (Cambridge University
Press)
\nonum Buck B and  Macaulay V A 1991 \textit{Maximum Entropy in Action} ( New York : Oxford University Press) 

\bibitem{sch}  Frieden B R 1989 {\it Am. J. Phys.} {\bf 57} 1004
\nonum Reginatto M 1998 {\it Phys. Rev. A} {\bf 58}  1775 
 
\bibitem{fish} Parwani R 2005 {\it J. Phys. A:Math. Gen.} \textbf{38} 6231 
 
 
\bibitem{RP1} Parwani R 2005 {\it  Ann. Phys.} \textbf{315} 419 

\bibitem {wk} Ng W K and Parwani R 2011 {\it Mod. Phys. Lett. A} {\bf 26} 681



\bibitem{nguyen} Nguyen L H and Parwani R 2009 {\it Gen. Rel. Grav.} {\bf 41} 2543
\nonum  Nguyen L H and Parwani R 2009 {\it AIP Conf.Proc.} 1115 p180 





\bibitem{siti}  Tarih S N and Parwani R, Preprint gr-qc/1107.3347v2 

\bibitem{meer}  Ashwinkumar M  and Parwani R, Preprint in preparation. 

\end{thebibliography}
\end{document}